\documentclass[aps,prd]{revtex4}
\pdfoutput=1

\usepackage[caption=false]{subfig}
\usepackage{amsmath,amssymb}
\usepackage{hyperref}
\usepackage{graphicx}
\usepackage{cleveref}

\newcommand{\lya}{Ly$\alpha$\ }
\newcommand{\msun}{M_{\odot}}
\newcommand{\kms}{\, {\rm km}\, {\rm s}^{-1}}

\begin{document}
\title{\boldmath Bosonic dark matter halos: excited states and relaxation
 in the potential of the ground state}

\author{Jorge Vicens}
\email{j.a.vicens.g@gmail.com}
\affiliation{Institut de Ci\`encies del Cosmos, Universitat de Barcelona (IEEC-UB), 08028 Barcelona, Catalonia, Spain}

\author{Jordi Salvado}
\email{jsalvado@icc.ub.edu}
\affiliation{Institut de Ci\`encies del Cosmos, Universitat de Barcelona (IEEC-UB), 08028 Barcelona, Catalonia, Spain}
\affiliation{Instituto de F\'isica Corpuscular, Val\`encia, Spain}

\author{Jordi Miralda-Escud\'e}
\email{miralda@icc.ub.edu}
\affiliation{Institut de Ci\`encies del Cosmos, Universitat de Barcelona (IEEC-UB), 08028 Barcelona, Catalonia, Spain}
\affiliation{Instituci\'o Catalana de Recerca i Estudis Avan\c cats,  08010 Barcelona, Catalonia, Spain}

\begin{abstract}
An ultra-light axion field with mass $\sim 10^{-22}\ {\rm eV}$, also known as
wave or fuzzy dark matter, has been proposed as a component of the dark matter
in the Universe. We study the evolution of the axion dark matter distribution
in the central region of a halo, assuming the mass is dominated by this axion
field, and that gravity is the only important interaction. We calculate the
excited axion states in the spherical gravitational potential generated by the
self-gravitating ground-state, also known as soliton. These excited states are
similar to the states of the hydrogen atom with quantum numbers $(n,l,m)$, here
designating oscillation modes of a classical wave. At fixed $n$, the modes with
highest $l$ have the lowest energy because of the extended mass distribution
generating the potential. We use an approximate analytical treatment to derive
the distribution of mass in these states when a steady-state is reached by
dynamical relaxation, and find that a corona with a mass density profile
$\rho \propto r^{-5/3}$ should be set up around the central soliton, analogous
to the Bahcall-Wolf cusp predicted for the stellar distribution around a
central black hole. The central soliton accretes dark matter from the corona
as dynamical relaxation proceeds and negative orbital energy flows out. This
density profile should remain valid out to the radius where the mass in the
corona is comparable to the mass of the central soliton; further than that, the
gravitational potential depends on the initial distribution of dark matter and
the relaxation time increases rapidly with radius.
\end{abstract}

\maketitle

\section{Introduction}
\label{sec:intro}

 ~\par The nature of dark matter remains one of the most important mysteries
in astrophysics and cosmology. The Cold Dark Matter theory (CDM) is
able to explain a wide range of observations: the power spectra of the
Cosmic Microwave Background fluctuations and large-scale structure
distribution of galaxies, the mass distribution in halos of galaxies and
clusters that is inferred from velocity distributions
\citep{Persic:1995ru}, X-ray
temperatures and gravitational lensing, the total mass density of the
Universe and the baryon fraction, the \lya forest power spectrum, etc.
\citep{Blumenthal1984,Frenk2012}.
At the same time, observations on the small-scale distribution of dark
matter have indicated possible discrepancies with the predictions
of CDM \cite{Weinberg:2013aya,Bullock2017,Primack:2009jr}, among them
the presence of flat cores in the mass density
profiles of dwarf galaxies dominated by dark matter
\cite{Dubinski:1991bm,Navarro:1997gj}.
At present there is an on-going debate to understand
if these cores, and other possible problems related to the abundance of
dwarf satellites \cite{McConnachie:2012vd,Klypin:1999uc,Moore:1999nt},
can be the result of baryonic processes (involving the
formation of stars in the central parts of low-mass dark matter halos,
formation of bars and exchange of angular momentum, and the ejection
of most baryons in galaxy winds), or if there is an
actual discrepancy indicative of a fundamental difference of
dark matter from the predictions of CDM.

  Several possible alternatives to CDM have been discussed which can
modify the predictions on the small-scale structure. For example, Warm
Dark Matter proposes that the dark matter particle has just the mass
required to give it a primordial velocity dispersion at present that is
comparable to the dispersions of dwarf galaxies. This would reduce the
abundance of dwarf galaxies and possibly predict cored distributions
instead of central cusps
\cite{Viel:2005qj,Colin:2000dn,Bode:2000gq,Lopez-Honorez:2017csg}.
Some of the best lower limits on the CDM particle mass have been obtained
from analysis of the \lya forest power spectrum \citep[e.g.,][and references therein]{Baur:2015jsy}.

  Ultralight axion dark matter is a different hypothesis first proposed
by \cite{Hu:2000ke,Goodman:2000tg,Peebles:2000yy} postulating that the
dark matter (or at least part of it) is a scalar or pseudo-scalar field
with a very small mass term, implying characteristic wavelengths on
galactic scales. The basic idea is that the dark matter
is a wave that obeys the equation of a scalar field $\psi$ evolving in
a gravitational potential $V$. The gravitational potential is at the
same time determined by the distribution of mass, which is dominated by
the dark matter itself except when baryons become more important in
central regions of halos due to dissipative processes. To understand the
behavior of these dark matter waves, it is generally useful to find
eigenstate solutions of the time evolution equation of the field.
In the non-relativistic limit (which we generally assume here as an
excellent approximation to any galactic dynamic situation far from
black hole horizons), the equation for eigenstates of $\psi$ is
\begin{equation}
\label{eq:wave1}
 - {\lambda_a^2\over 2}\, \nabla^2 \psi + {V\over c^2}\psi =
 \frac{i \lambda_a}{c} {\partial \psi \over \partial t} = 
 {\epsilon \over c^2}\, \psi ~,
\end{equation} 
where $\epsilon$ is an eigenvalue that is the energy per unit mass of
the field, and $\lambda_a=\hbar/(m_a c)$ is the reduced Compton wavelength,
related to the particle mass $m_a$ of the scalar field. This is the
Schr\"odinger equation for a particle of mass $m_a$ moving in a
potential $V$. When expressed in
terms of the wavelength $\lambda_a$, equation \ref{eq:wave1} describes
the evolution of the field $\psi$ as a purely classical wave, and there
is no reference to the Planck constant. The relation
to quantum physics appears only if we insist in relating the field to
single particles of mass $m_a$.
In astrophysical situations,
the number of particles per stationary state, or occupation number, is
extremely large, and the field $\psi$ behaves according to classical
physics. However, many of the concepts and terminology from quantum
mechanics that are familiar from the study of atomic physics are often
useful for describing the phenomenology that this equation gives rise
to.

  When the only interaction that this scalar field is subject to is
gravity, and we assume that this scalar dark matter is the only
important mass that is present (neglecting any contribution from baryons
or other components), the gravitational potential is given by the
Poisson equation with a mass density generated by the field,
\begin{equation}
\label{eq:Poisson}
\nabla^2V(r)=4\pi G M |\psi(r)|^2 ~,
\end{equation}
where the integral of $|\psi|^2$ over all volume is normalized to unity,
and $M$ is the total mass of the system. The combination of these two
equations is referred to as the {\it Schr\"odinger-Poisson} equation.

 When looking for eigenstates of these equations, one finds first the
solution of the ground state, in which a total mass $M$ of axions forms
a self-gravitating object with the lowest possible energy. This solution
was first found in \cite{Ruffini:1969qy}, giving it the name of a
{\it boson star}. The name was not very appropriate, because stars in
astrophysics do not refer to any kind of self-gravitating objects, but
to baryonic ones that have produced energy in their interiors through
nuclear fusion reactions (brown dwarfs can be considered stars if we
include the small amount of initial deuterium burning in these
reactions). At the same time, numerical simulations of the evolution
of a scalar field \cite{Schive:2014dra,Schive:2014hza} that directly
solve the Schr\"odinger-Poisson equation with numerical techniques have
identified the formation of collapsed objects and designated them as
{\it solitons} \cite{Marsh:2015wka}. This has perhaps obscured the fact
that the collapsed,
stationary objects that form when a random initial scalar field is
numerically evolved are the same solution that was found by
\cite{Ruffini:1969qy}, except that they are surrounded by a dark matter
halo that is not part of the stationary
solution. In this ground state solution, a mass $M$ of the
axion field is confined to a radius $R$ corresponding to a virialized
halo with velocity dispersion $v^2\sim GM/R$, which is equal to the de
Broglie wavelength $R\sim \lambda_a c/v \sim \lambda_a c (R/GM)^{1/2}$,
implying that $R\sim \lambda_a^2/R_g$, where $R_g=GM/c^2$ is half the
Schwarzschild radius of the mass $M$ \cite{Lin:2018whl}. At fixed mass $M$, the object
cannot collapse to a smaller radius than this ground state because the
increased kinetic energy required to further confine the scalar wave is
more than the gravitational energy obtained from the collapse. We shall
also refer to these equilibrium self-gravitating objects as scalar
field {\it solitons}.

  Apart from the ground state or soliton solution, one can find other stationary
states of the coupled Schr\"odinger-Poisson equation, representing
excited states that are also stationary. These states can be easily
computed if spherical symmetry is imposed, and have been presented
previously (e.g., \cite{Bernstein:1998,Hui:2016ltb}). These excited states are,
however, not physically very meaningful: they represent a solution in
which all the axions (or all the energy of the classical axion field)
are in an excited state, and the ground state is empty. In practice
there would never be any reason why the ground state should be empty.
In an evolving classical situation, any small time variation of the
potential due to a small part of the axion field being in different
states would cause transitions from one eigenstate to all others.
If we consider the quantum mechanical case with individual axions, even
if all the axions are initially placed in one excited state, there would
always be a spontaneous decay rate, however small, to lower energy
states, and stimulated decay would subsequently increase exponentially
with time as more axions occupy other states.

  In this paper we consider a different set of eigenstates that are of
greater use to understand the evolution of the classical scalar field
under a broad range of conditions: we assume that most of the mass of
the axion field is in the ground state, and that a small fraction is in
other excited states. We then approximate the gravitational potential V
as being the one created by the axion mass distribution in the ground
state, and calculate the excited eigenstates in this fixed potential.
The small contribution to the gravitational potential from the mass in
the excited states can then be considered as a perturbation that causes
small transition rates among the various excited states. These
transition rates can be considered as a perturbation expansion for the
evolution of a classical field, but also correspond in the usual quantum
language to the gravitational scattering of two axions from two incoming
to two outgoing states.

 The phenomenology of ultra light axion dark matter has been studied
in the context of large-scale precision cosmology
\cite{Hlozek:2014lca}, using astrophysical probes such as the \lya
forest \cite{Irsic:2017yje}, pulsars \cite{Khmelnitsky:2013lxt, deMartino:2018krg,
  Blas:2016ddr}, gravitational waves \cite{Dev:2016hxv}, and
laboratory based experiments
\cite{Aoki:2016kwl,Brdar:2017kbt,Krnjaic:2017zlz}. Although some
constraints can be set in some parts of the parameter space, a
conclusive answer has yet to be found. In particular, models in which
ultralight axions constitute only a fraction of the dark matter, with
other components providing small-scale power, can probably circumvent
the difficulties in trying to satisfy observational constraints from
dwarf galaxy mass profiles and the \lya forest at the same time.

  We present in section \ref{sec:sol} the solutions to these excited
states. Then, in section \ref{sec:rel} we describe an application for
the relaxation process of axion dark matter in a spherical distribution
around the ground state, and we give an approximate solution to the way
these excited states are populated when this relaxation occurs.
We conclude in section \ref{sec:con} discussing this application as a
prediction for the dark matter distribution in various galaxies in the
axion dark matter model, and future directions for improving this
calculation.

\section{Excited states in the potential of the bosonic ground state}
\label{sec:sol}

\subsection{The ground state solution}

 Following the mathematical treatment in \cite{Bernstein:1998}, and
similarly in \cite{Ruffini:1969qy}, we apply a re-scaling of the
physical variables of our equations to make them dimensionless,
\begin{eqnarray}
  \label{eq:scale1}
  r & \rightarrow & \left(\frac{\hbar^2}{G M m_a^2}\right) =
 {\lambda_a^2\over R_g} r ~, \\
  \label{eq:scale2}
  \psi &\rightarrow & \left( \frac{\hbar^{2}}{G M m_a^2}
                      \right)^{-\frac{3}{2}} \psi  =
 \left( {\lambda_a^2\over R_g} \right)^{-3/2} \psi ~ ,\\
  \label{eq:scale3}
  V & \rightarrow & \left(\frac{G^2M^2m_a^2}{\hbar^2}\right) V =
 {c^2R_g^2 \over \lambda_a^2} V ~ ,\\
  \label{eq:scale4}
  \epsilon & \rightarrow & \frac{c^2R_g^2}{\lambda_a^2} \epsilon ~,
\end{eqnarray}
where $R_g = GM/c^2$.
With this transformation, equations
\ref{eq:wave1} and \ref{eq:Poisson} can be written as,
\begin{equation}
\label{eq:wave_sph}
\left( {1\over 2} \nabla^2  - V_\alpha \right)\psi_\alpha =
 \epsilon_\alpha \psi_\alpha  ~,
\end{equation}
and
\begin{equation}
\label{eq:PoissonRho}
\nabla^2 V = 4\pi |\psi_\alpha|^2 ~,
\end{equation}
where the subindex $\alpha$ labels the eigenstate solutions with
eigenvalues $\epsilon_\alpha$.

\begin{figure*}[ht]
  \subfloat[\label{fig:subim1} Radial Wave Function]{%
    \includegraphics[width=0.5\linewidth, height=5cm]{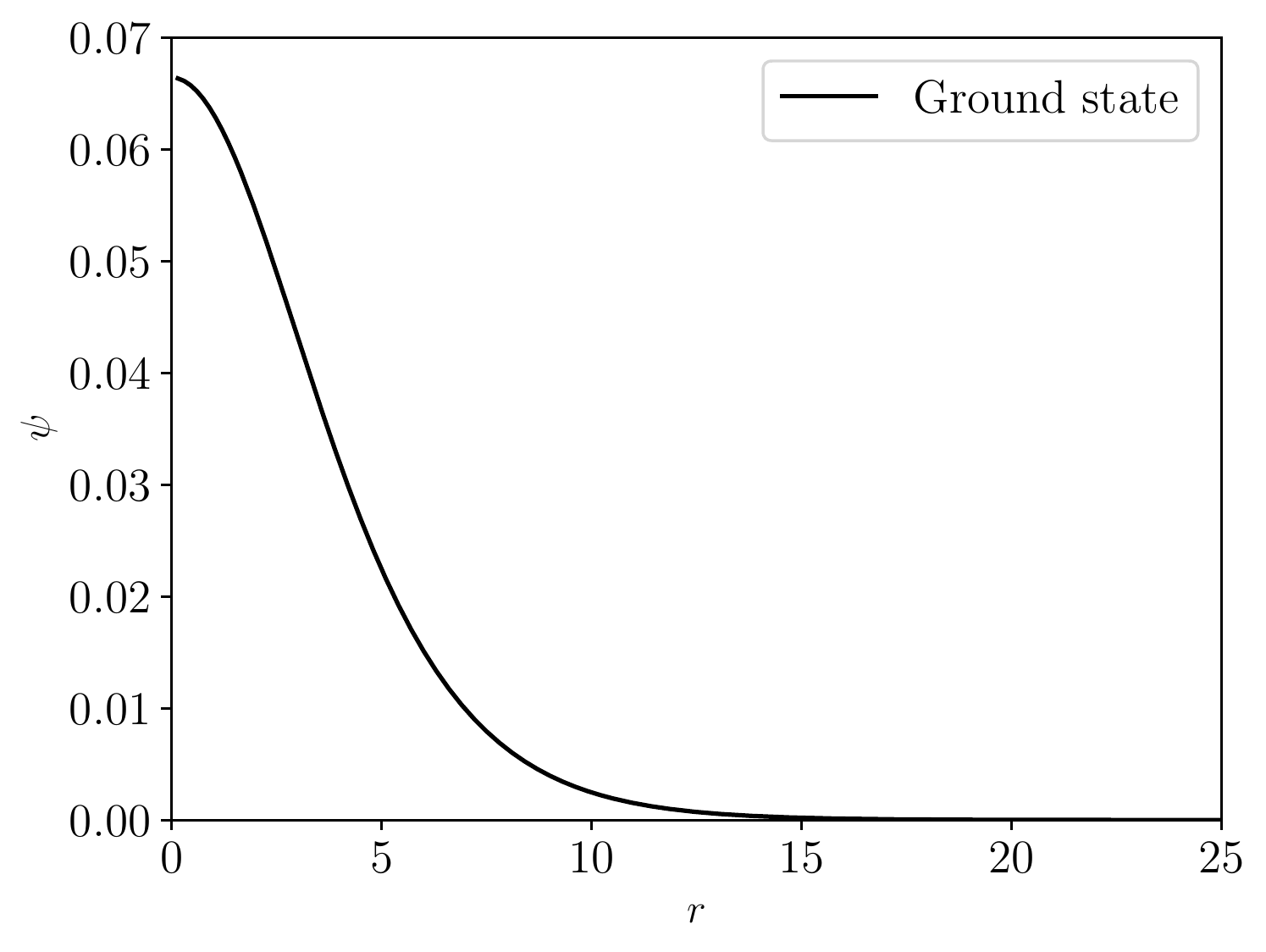}
  }  \subfloat[\label{fig:subim2} Gravitational Potential]{%
    \includegraphics[width=0.5\linewidth, height=5cm]{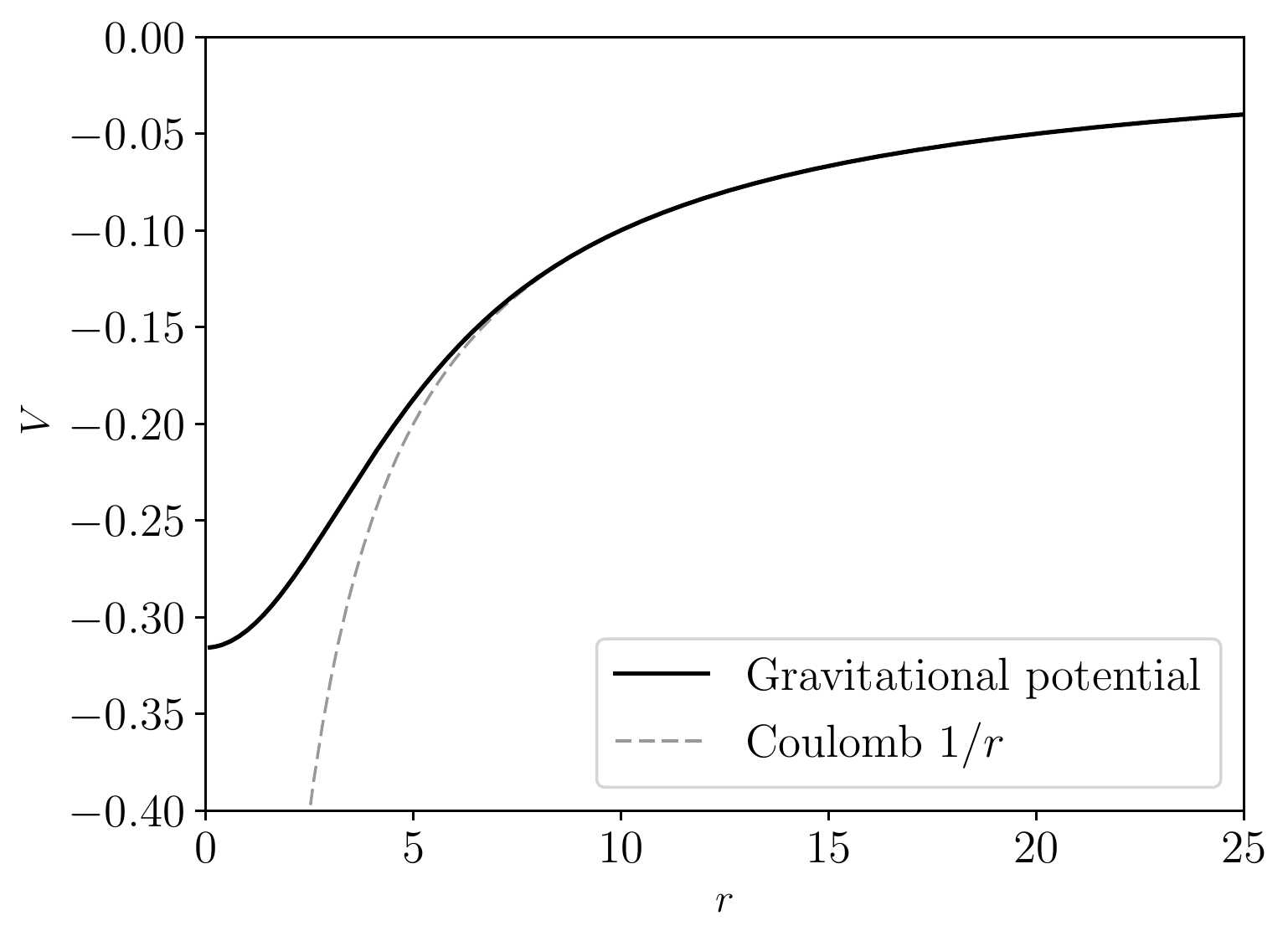}
  }
  \caption{\label{fig:image2} Numerical solutions for self-gravitating ground state wave
    function and gravitational potential. The result is shown in terms of the
    dimensionless magnitudes shown in equations \ref{eq:scale1} to
    \ref{eq:scale4}. }
\end{figure*}

  The ground state solution $\psi_1$ is spherically symmetric
\cite{Moroz:1998dh}, so it is solved by replacing
$\nabla^2\psi_1= (1/r) \partial^2/\partial r^2(r\psi_1)$. We use the
numerical method described in \cite{Bernstein:1998} to convert the
differential equations to second-order finite difference equations
with two-point boundary conditions requiring the wave function to go
to zero at infinity and its derivative to go to zero at the origin. We
reproduce the solution found by \cite{Ruffini:1969qy}
in figure \ref{fig:subim1} and \ref{fig:subim2}, showing
the ground state wave function and the corresponding potential in
terms of the dimensionless quantities in equations
\ref{eq:scale1} to \ref{eq:scale4}.

  In the second plot, the potential of a point mass with the same mass
$M$ as the bosonic ground state is shown for comparison. The solution
for the potential of the point mass is of course the ground state of the
non-relativistic hydrogen atom. The only difference in the
self-gravitating bosonic object is that the mass is extended, so the
potential is less deep in the center and the wave solution is also
more extended compared to the hydrogen atom solution.

\subsection{Excited states and energy spectrum}

  We now compute the stationary solutions for the excited states assuming
that the gravitational potential is dominated by the ground state, which
we designate as $V_1(r)$ and is fixed. The self-interaction among the
excited states is neglected, so can also solve for the states that are
not spherically symmetric preserving the spherical symmetry of the
equation. The solutions are, as known from atomic physics, the spherical
harmonic functions $Y_{lm}$, times a radial function that obeys the
equation
\begin{equation}
\label{eq:Schrodinger_angular}
 {1\over r} { \partial^2 [r \psi_{nl}(r) ] \over \partial r^2 } +
 \left( V_1(r)+ \frac{l(l+1)}{2m_ar^2}\right) \psi_{nl}(r) =
\epsilon_{nl} \psi_{nl}(r),   
\end{equation}
where $\epsilon_{nl}$ is the energy eigenvalue of excited state $n$
with angular momentum number $l$, and $V_1(r)$ is the solution for
the ground state potential shown in figure \ref{fig:subim2}

\begin{figure}[ht]
\includegraphics[width=1\linewidth]{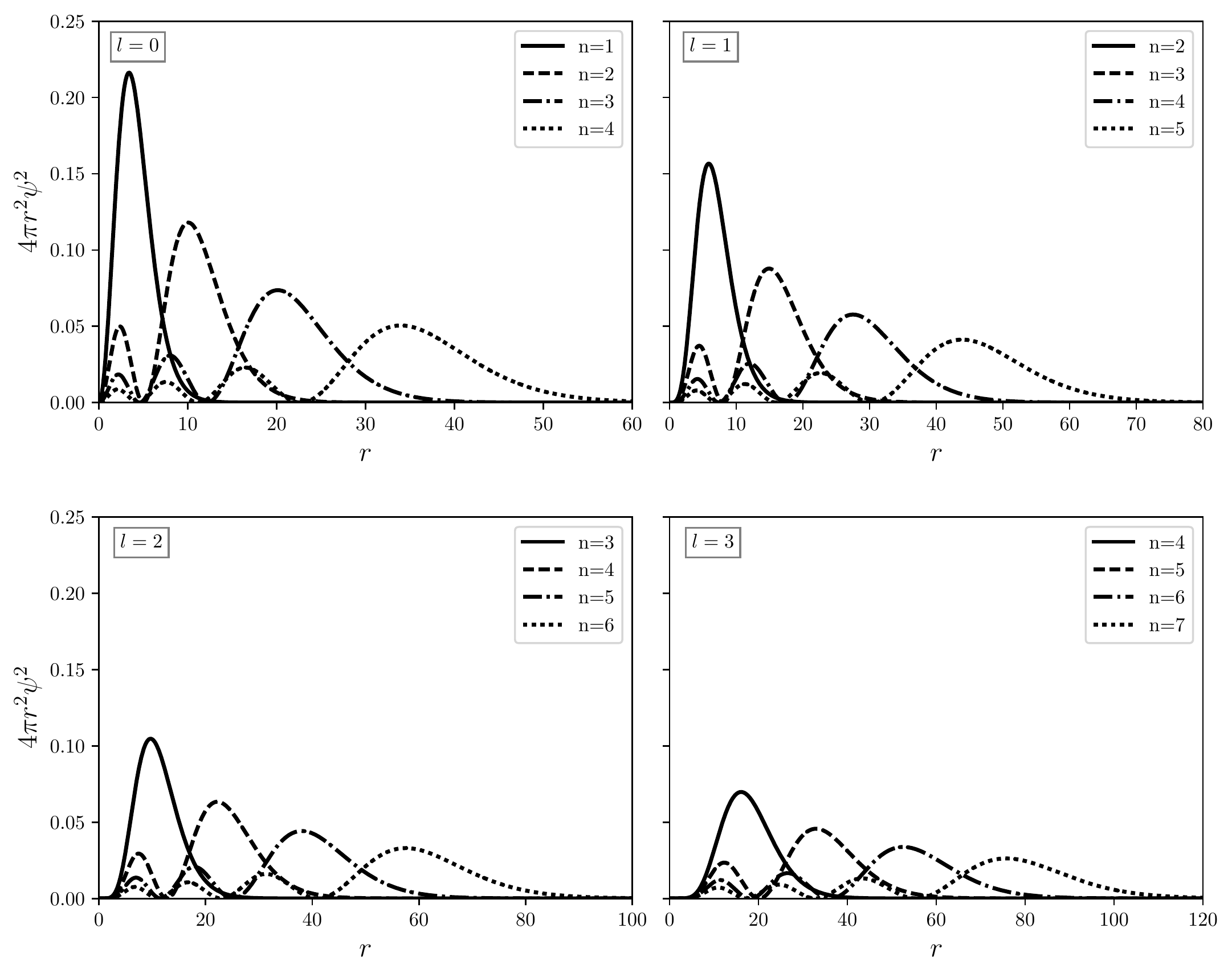} 
\caption{Numerical solutions for several values of $(n,l)$ for the
normalized probability per unit $r$, $4\pi r^2\psi^2$, plotted in terms
of dimensionless radius.}
\label{fig:ExcitedStates}
\end{figure}

  The wave function numerical solutions for several values of $(n,l)$
are shown in \cref{fig:ExcitedStates}. As the principal quantum number
$n$ increases, the wave function is more extended with a characteristic
half-radius $r \sim n^2$, and the potential of the mass of the ground
state is increasingly well approximated as that of a point mass. The
eigenfunctions therefore become closer to the hydrogen atom solutions as
$n$ increases.

\begin{figure}[ht]
  \includegraphics[width=\textwidth]{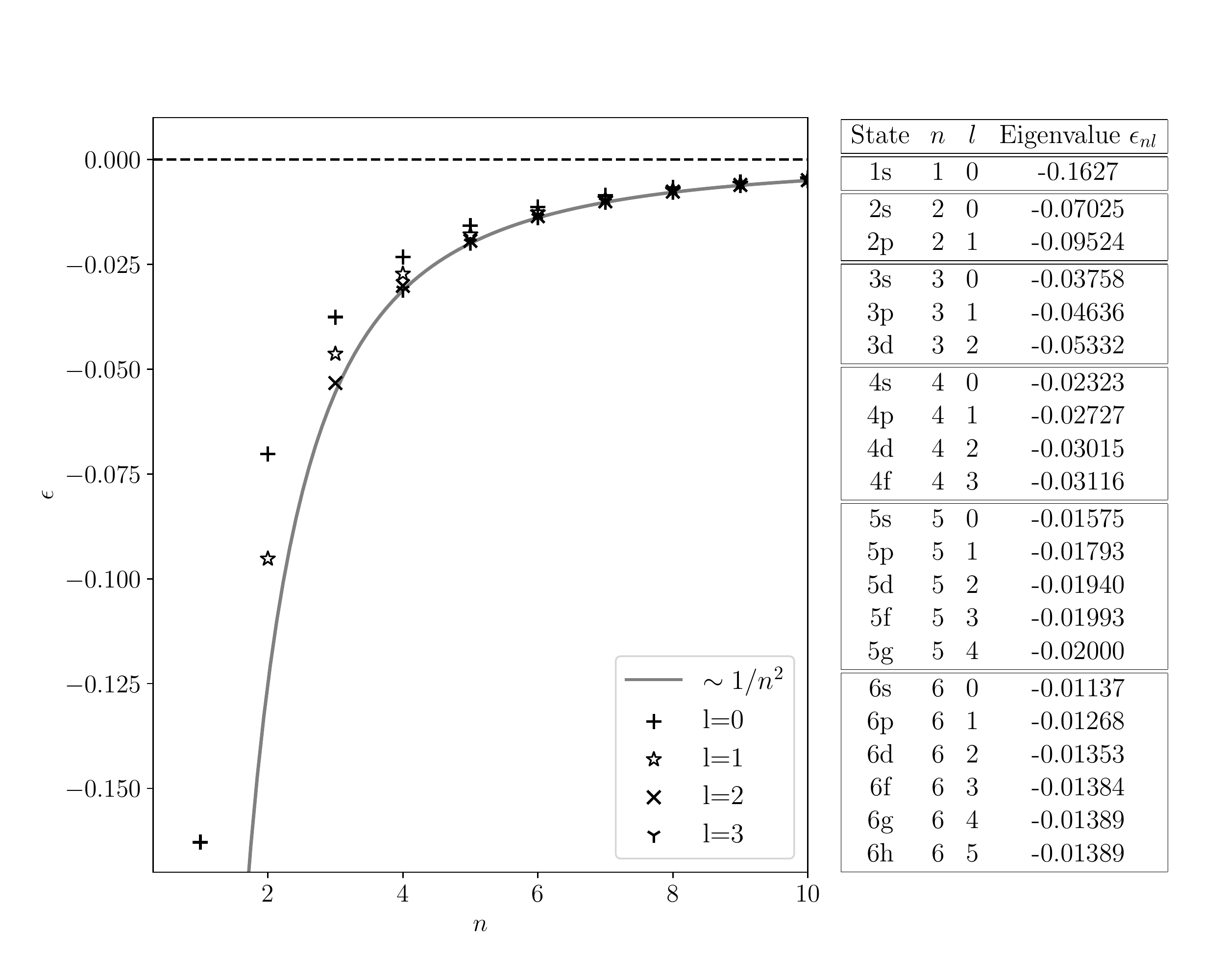}
  \caption{Energy spectrum of excited states in the potential of the
    ground state for $n$ up to 10 and all the values $0<l<n-1$.}
  \label{fig:spectrum}
\end{figure}

  Figure \ref{fig:spectrum} shows the energy eigenvalues that solve the
Schr\"odinger-Poisson equation with the potential $V_1(r)$. The
spherically symmetric eigenstates, with $l=0$, are shown with straight
crosses, and the solid line shows the hydrogen atom solution
$\epsilon \sim 1/n^2$ for comparison. The spatial extent of the ground
state mass flattens the potential and decreases the energy eigenvalues
in absolute value, an effect that becomes smaller with growing $n$.
States with
angular momentum avoid the central region, and therefore are less
affected by the central flattening of the potential and have energies
closer to that of the hydrogen atom. The energies $\epsilon_{nl}$ increase
with $l$ in absolute value at fixed $n$, contrary to the effect of
relativistic corrections in the hydrogen atom (which make the potential
steeper than Keplerian, whereas the extended mass makes the potential
shallower than Keplerian). Consequently, the lowest energy states in
bosonic self-gravitating systems at fixed $n$ are the ones with largest
angular momentum. At high $n$ the eigenvalues of different $l$ are
increasingly degenerate as the potential approaches that of a point mass.

\section{Evolution of axion dark matter halos under dynamical relaxation}
\label{sec:rel}

 ~\par Having established the form of the excited states of bosonic dark
matter and their energy spectrum when the gravitational potential is
dominated by the ground state, we now consider the way these states are
populated when the dark matter is in the process of dynamically
relaxing.

\subsection{Relaxation timescale in halos
containing axion dark matter}

  The process of dynamical relaxation can be understood classically as
the result of a perturbation on a smooth and time-independent
gravitational potential. The scalar field state can be expressed as
$\psi = \psi_1 + \psi_e$, where $\psi_1$ is the ground state and
$\psi_e$ is the small contribution of the axions in all the excited
states. The gravitational potential has a first order perturbation term
in the Laplacian appearing in the Poisson equation that is proportional
to $\psi_1 \psi_e^{*} + \psi_1^{*} \psi_e$, affecting the evolution of
the field. This effect can be expressed as a collisional term in a
Boltzmann equation for the evolution of the scalar field amplitude in
all the states, where the gravitational interaction of the field in two
excited states causes a transition into the ground state and a third
excited state, and viceversa. In quantum language, this is understood as
the effect of stimulated gravitational scattering of two axions in
excited states, into an axion in the ground state and an axion in an
excited state that acquires the energy of the two incoming excited
axions. The exact quantum transition amplitudes also include the
spontaneous scattering rate, but this is negligible when the quantum
occupation numbers of all the states are very large, as is the case in
all realistic astrophysical situations involving ultralight axions. We
note that, in general, axion fields may be subject to self-interactions
other than gravity, but here we assume that self-gravity is the only
important interaction that induces dynamical relaxation.

  Again, we find here that even though the problem we are addressing is
one of purely classical waves, the quantum language that has been
developed in atomic physics is conceptually useful. The precise method
to compute transition rates among excited states induced by this
gravitational scattering is to calculate transition amplitudes using
Fermi's golden rule, for the process of gravitational scattering of two
incoming to two outgoing axions. In this paper, however, we will not
carry out this detailed calculation, which we leave for later work.
Instead, we limit this work to deriving simple scaling relations for the
relaxation time and the implied mass distribution in the excited states.
For this purpose, we use the simple rule that the scalar field
dynamically relaxes as a collection of quasi-particles that represent
the number of independent oscillation modes of the scalar field that
exist on a given radial shell around the soliton \citep{Hui:2016ltb}.
In terms of the excited
states we have calculated, and for states with $n \gg 1$, the
characteristic radial extent of the states is $r\sim n^2$ (like in the
hydrogen atom), and the number of different states up to quantum number
$n$ scales as $N_q\sim n^3$, so the number of available states within
$r$ scales as $N_q\sim r^{3/2}$. This corresponds to the scaling of the
phase-space volume within $r$ of bound orbits in a Keplerian potential,
in which the velocity dispersion is $v\sim (GM/r)^{1/2}$ and
$(vr)^3 \sim (GMr)^{3/2}$.

  Classically, the relaxation time calculated from gravitational
scatterings of a system of particles of a fixed mass $m$ and mass
density $\rho$ with a velocity dispersion $v$ is \citep{Binney87}:
\begin{equation}
  t_{\rm rel} = 0.34 {v^3\over G^2 m \rho \log\Lambda} ~,
\end{equation}
where $\log\Lambda$ is the Coulomb logarithm factor. If the total mass
of the system within $r$ that generates the gravitational potential is
$M$, this can be expressed as
\begin{equation}
\label{eq:relax}
  t_{\rm rel} \simeq {t_{\rm orb}\over \log\Lambda} {M\over N_p m}{M\over m} ~.
\end{equation}
where $N_p$ is the number of particles within $r$ and
$t_{\rm orb}= r/v$ is the orbital time. This has a simple
interpretation: when the particles account for all the mass of the
system $M$, then the relaxation time scales as $N_p t_{\rm orb}$, and
when the particles are only a fraction $f_p=N_p m/M$ of the mass of the
system, the relaxation time (at fixed $M/m$) is increased by the
inverse of this fraction because there are fewer particles of mass $m$
available to produce scatterings.

  This equation also determines the relaxation time of the scalar field
in the potential of a central mass $M$ as a function of the number of
quasi-particles, except that now the quasi-particle mass $m$ is not
constant but varies with radius depending on the scalar density.
The number of quasi-particles, on the other hand, is fixed to the number
of excited states of the field available at each radius $r$.

\subsection{The radial profile of relaxing axion dark matter around the
 ground state}

  Let us now recall the derivation of the equilibrium distribution of
particles orbiting in a Keplerian potential after they have dynamically
relaxed. This problem was addressed to understand the evolution of a star
cluster moving in the potential of a central massive black hole, which
was predicted to reach a stationary distribution known as the
Bahcall-Wolf cusp \citep{Bahcall76,Bahcall77}. We are interested here
only in the radial profile shape, so we ignore the slowly varying
Coulomb logarithm term. One can first try the assumption that relaxation
leads to a flow of particles toward the center of the potential that is
constant in radius \citep{Peebles1972}, if stars are simply assumed to
disappear once they reach the center and are destroyed by the black
hole. In this case we would require $N_p/t_{\rm rel}$ to be constant,
corresponding to the assumption that the $N_p$ particles at radius $r$
move to an interior shell after time $t_{\rm rel}$ and that this flow
has to be conserved at all radii. This implies (using
$t_{\rm orb}\sim r^{3/2}$ in a Kepler potential)
\begin{equation}
\label{eq:rho0}
 N_p m/t_{\rm rel} \sim N_p^2 m^3/M^2/r^{3/2} \sim {\rm constant} ~, \qquad
 \rho(r) \sim N_p m/r^3 \sim r^{-9/4} ~.
\end{equation}
This density profile assumes that the mass flow is locally determined by
the relaxation time at each radius. However, the orbital energy of the
particles also needs to be conserved and has to flow out radially as
particles move in. The steep density profile in \ref{eq:rho0} does not
allow for the orbital energy to flow out fast enough, because the
negative energy per particle decreases as $1/r$ in the Keplerian
potential. The mass flow is therefore determined by the relaxation time
at an outer radius that fixes the rate at which orbital energy can flow
out. A constant transfer rate of orbital energy is then the correct
assumption, implying
\begin{equation}
\label{eq:rho0b}
 N_p m/(rt_{\rm rel}) \sim N_p^2 m^3/M^2/r^{5/2} \sim {\rm constant} ~,
 \qquad \rho(r) \sim N_p m/r^3 \sim r^{-7/4} ~.
\end{equation}

  We now repeat the same derivation for the system of relaxing
quasi-particles of the scalar field. In this case, the scalar wave is
exchanging energy among modes as relaxation proceeds, leading to an
accretion rate onto the ground state while some of the scalar field is
excited to higher energy levels. The assumption of a constant
mass flow is again not self-consistent, because even though it would
allow for conservation of the scalar field mass as it accretes in a
steady-state solution toward the central soliton, it does not provide
for a way to absorb the increasing negative potential energy of the
scalar field as it moves inward. The scalar field loses orbital energy
by scattering some of the axions into higher energy states at larger
radius, but just like in the case of classical particles, this occurs
only at the relaxation rate. The majority of axions are not scattered
into nearly empty, unbound states of axions, but into other populated
bound states that also need to transfer their potential energy outwards
in radius, so the solution requires a constant outward flow of orbital
energy. For quasi-particles, however, their effective mass $m_q$ depends
on radius and the number of quasi-particles is determined to evolve
radially as $N_q\sim r^{3/2}$, which implies
\begin{equation}
 N_q m_q/(rt_{\rm rel}) \sim N_q^2 m_q^3/M^2/r^{5/2} \sim
 r^{1/2} m_q^3/M^2 \sim {\rm constant} ~,
\end{equation}
\begin{equation}
\label{eq:rho0c}
 \rho(r) \sim N_q m_q/r^3 \sim r^{-5/3} ~.
\end{equation}

  In summary, we find that the relaxation of scalar field dark matter
in a Keplerian potential determined by the mass of the central soliton
should give rise to a corona of scalar dark matter surrounding the
soliton, with a mass density profile $\rho \sim r^{-5/3}$, determined by
a constant rate of orbital energy flowing out as the dark matter
accretes on the soliton. Our derivation neglects the influence of
particles or quasi-particles that are ejected after a gravitational
encounter at a sufficiently high velocity to become unbound, removing
energy from the dynamical system directly, but this should likely be a
small contribution because the total relaxation rate is enhanced by the
Coulomb logarithm factor.

\subsection{Consequences for density profiles in axion dark matter halos}

  When a halo made of axion dark matter has been in dynamical
equilibrium for a time longer than the central relaxation time $t_{\rm rel}$,
the dark matter should start accumulating in the central ground state
or soliton. The mass of the soliton should increase as more dark matter
relaxes, with some mass accreting from the excited states to the soliton
and other mass moving to higher energy states. The accretion occurs from
a corona of relaxing dark matter surrounding the soliton, with an
$r^{-5/3}$ density profile, extending out to a radius $r_c$ at which the
mass of the corona is similar to the mass of the soliton. Beyond this
radius the gravitational potential is no longer Keplerian, and the dark
matter density profile changes and depends on the initial conditions.

  The accretion rate of dark matter onto the soliton can be related to
the mass of the soliton $M_s$ and the outer radius $r_c$ of the
$r^{-5/3}$ corona as follows. The relaxation time at $r_c$ is, from
equation \ref{eq:relax},
\begin{equation}
 t_{\rm rel}(r_c) \simeq t_{\rm orb}(r_c) {2 M_s \over N_q(r_c) m_q(r_c)}\,
  {2 M_s \over m_q(r_c)} ~.
\end{equation}
We have assumed that the total mass within $r_c$ is
$M_s + M_c(r_c)\simeq 2M_s$. The excited states occupying the volume
within $r_c$ have quantum number $n_c \simeq (2r_c/r_s)^{1/2}$, where
$r_s\simeq 4 \lambda_a^2/R_g$ is the half-mass radius of the soliton
(see figures \ref{fig:image2} and \ref{fig:ExcitedStates} for the
approximate factors of 2 and 4 we have introduced). The number
of excited states, or quasi-particles, within $r_c$ is $N_q(r_c)\simeq
n_c^3/3\simeq (r_c/r_s)^{3/2}$, and the mass of each quasi-particle is
$m_q(r_c)\simeq M_s/N_q(r_c)$. The orbital time at $r_c$ can also be
expressed as
$t_{\rm orb}(r_c)\simeq 2^{-1} t_{\rm orb}(r_s)\, (r_c/r_s)^{3/2}$,
so we finally have
\begin{equation}
 t_{\rm rel}(r_c) \simeq {2\over \log\Lambda }\, t_{\rm orb}(r_s)
 \left( {r_c\over r_s}\right)^3 ~.
\end{equation}
The timescale required to accrete the mass $M_s$ onto the soliton is,
however, much longer than $t_{\rm rel}(r_c)$. The relaxation process at
$r_c$ allows the orbital energy contained in the quasi-particles at
$r_c$ to be transported outwards, but the orbital energy per unit mass
at $r_c$ is a factor $r_s/r_c$ smaller than at $r_s$. The timescale for
{\it accretion} of all the mass $M_s$ onto the soliton, $t_{\rm acc}$,
is therefore $r_c/r_s$ times longer than $t_{\rm rel}(r_c)$:
\begin{equation}
 t_{\rm acc}(r_c) \simeq {2\over \log\Lambda}\, t_{\rm orb}(r_s)
 \left( {r_c\over r_s}\right)^4 ~.
\end{equation}
We point out again that this accretion timescale assumes that the
negative orbital energy that needs to be transported outwards is
conducted by means of diffusive gravitational scatterings. Large
scatterings that can eject axion streams directly out into
unbound orbits are not included, and they may increase the rate of
accretion onto the soliton. This effect, however, is also important
for the classical stellar dynamics problem of the evolution of a cluster
of stars around a massive black hole.
  
  Inside the radius $r_c$, the relaxation time varies slowly as
$t_{\rm rel} \propto r^{1/3}$. However, outside $r_c$ the relaxation
time increases very rapidly with radius. For example, if the dark
matter density gives rise to a flat rotation curve as in many galaxies,
with a density profile $\rho(r)\propto r^{-2}$, then the orbital time
is $t_{\rm orb} \propto r$, the number of states is $N_q\propto r^3$,
and $t_{\rm rel}\propto t_{\rm orb} N_q \propto r^4$. The negative
orbital energy flowing out therefore accumulates near $r_c$, pushing
out the dark matter and flattening the profile at radius
just above $r_c$.

\begin{figure}[ht]
  \begin{center}
  \includegraphics[width=\linewidth]{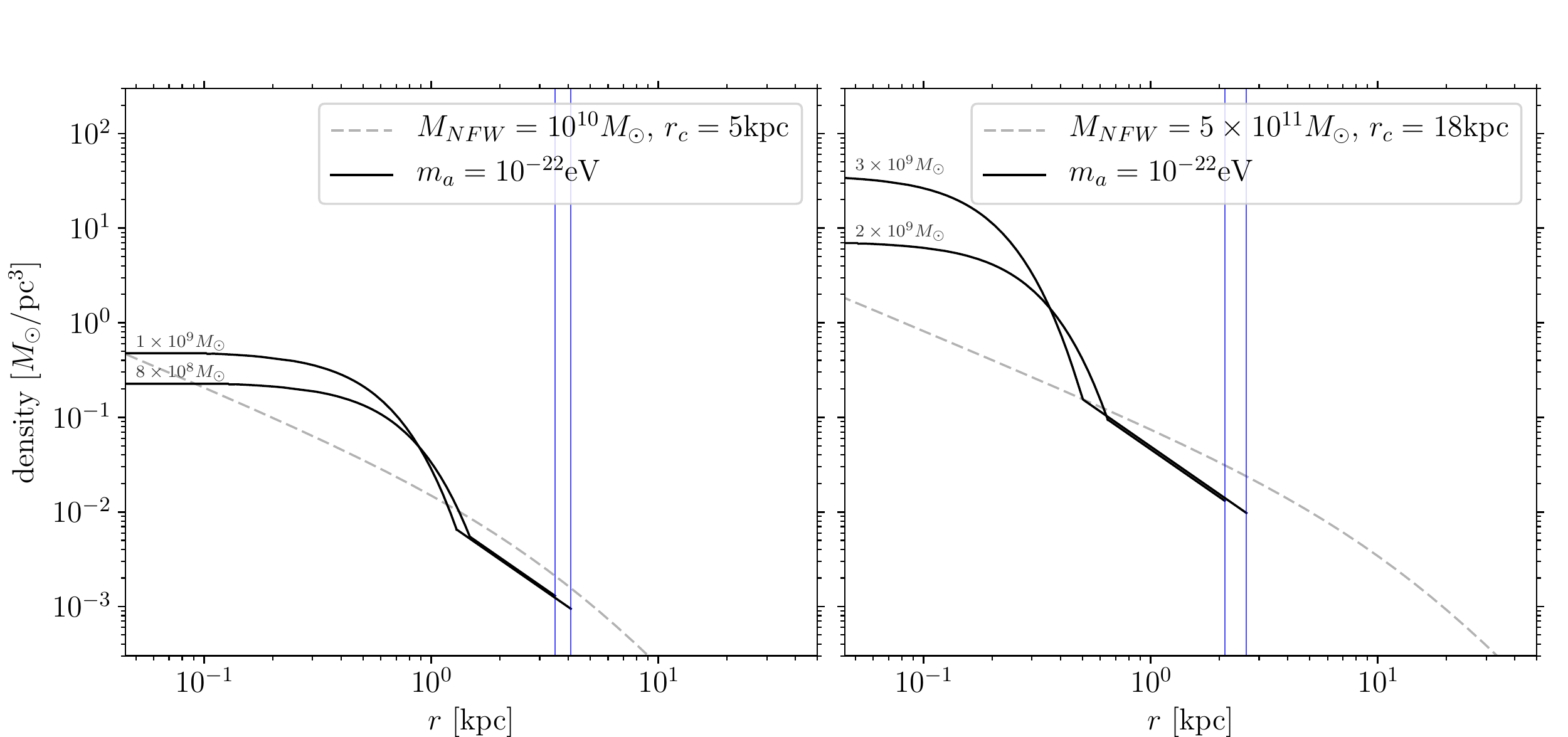}
  \label{fig:density}
  \caption{ {\it Light dashed lines}: Density profile of axion dark
matter halos of total mass $M_h=10^{10}M_\odot$ (left) and
$M_h=5\times 10^{11}M_\odot$ (right), with the NFW profile and indicated
core radii (the total mass is within $r_{max}=50 {\rm kpc}$ (left) and
$r_{max}=10 {\rm kpc}$ (right) ). {\it Solid lines:}
Density profile of the soliton and surrouding $r^{-5/3}$ corona, with
total masses $2M_s$ indicated in each curve, split in a mass $M_s$ in
the soliton and a mass $M_s$ in the corona. Blue vertical lines indicate
the radius where the initial NFW profile contains a mass equal to
$2M_s$, from which the mass of the soliton and corona have been
accreted.}

  \end{center}
\end{figure}

  To find the amount of mass $M_s$ that should have accreted on the
central soliton in a dark matter halo after relaxation has started in
the center, we solve for the radius $r_{\rm cor}$ that was initially (after the
formation of the halo) containing a total mass $2M_s$, and where the
accretion time $t_{\rm acc}(r_{\rm cor})$ is equal to the halo age. We assume
that there are no other mass components (either baryonic or other types
of dark matter) that affect the gravitational potential. As illustrating
examples, we consider an axion mass $m_a=10^{-22}\, {\rm eV}$ and dark
matter halos with the NFW profile
\cite{Navarro:1997gj},
\begin{equation}
 \rho(r)= {\rho_0 r_c^3\over r (r+r_c)^2 } ~,
\end{equation}
with a cusp radius $r_c=5 \,{\rm kpc}$ and mass
$M_{NFW}=10^{10} \msun$ within $r_{\rm max}=50{\rm kpc}$ (left panel of figure
\ref{fig:density}), and a cusp radius $r_c= 18 \, {\rm kpc}$ and mass
$M_{NFW}=5\times 10^{11} \msun$ within $r_{\rm max}=180{\rm kpc}$ (right panel in
the same figure). The first example represents a dwarf galaxy halo
similar to the Large Magellanic Cloud, and the second a galactic halo
like that of the Milky Way. Two possible profiles of the axionic soliton
and the surrounding $r^{-5/3}$ corona are shown in both panels, with
their total mass $2M_s$ indicated next to each curve. The lowest mass
soliton in the left panel, with $M_s=4\times 10^8 \msun$, has a
half-mass radius for the soliton $r_s\simeq 4\lambda_a^2/R_g\simeq
0.85\, {\rm kpc}$, where the circular velocity is $v_c(r_s)\simeq
32 \kms$ and the orbital time is
$t_{\rm orb}(r_s)\simeq 2.6\times 10^7\, {\rm yr}$. The mass
$2M_s=8\times 10^8\msun$ is contained in the initial NFW profile inside
a radius $r_{\rm cor}\simeq 4.5\, {\rm kpc}$ (indicated by the left,
blue vertical line), from which the accretion time is
$t_{\rm acc}(r_{\rm cor})\sim (r_{\rm cor}/r_s)^4 t_{\rm orb}(r_s)
\simeq 2\times 10^{10}\, {\rm yr}$. Therefore, the mass
$M_s=4\times 10^8 \msun$ is roughly the mass that would have been
accreted on the soliton at the present age of the Universe. The rate of
mass accretion decreases sharply with time: for $M_s=5\times 10^8\msun$,
the accretion time at $r_{\rm cor}\simeq 5.3\, {\rm kpc}$ increases to 
$t_{\rm acc}(r_{\rm cor})\sim 6\times 10^{10}\, {\rm yr}$.

  For a halo mass $M_{NFW}=5\times 10^{11}\msun$ (right panel), we find
that for $M_s=10^9\msun$, the accretion time at the radius
$r_{\rm cor}=1.8\, {\rm kpc}$ that contains an initial mass $2M_s$ is
$t_{\rm acc}(r_{\rm cor})\simeq 3\times 10^9\, {\rm yr}$, and for
$M_s=1.4\times 10^9\msun$, we find $r_{\rm cor}=2.2\, {\rm kpc}$ and
$t_{\rm acc}(r_{\rm cor})\simeq 1.4\times 10^{10}\, {\rm yr}$. The mass
of the soliton increases very slowly with the halo mass and the age,
and is strongly dependent on the axion mass $m_a$ and on the presence
of baryons or other dark matter components altering the gravitational
potential.

\section{Discussion and conclusions}
\label{sec:con}

  We have presented a numerical calculation of the excited states of
a scalar field in the gravitational potential of the mass distribution
of the self-gravitating ground state. This solution for the excited
states, which approaches the hydrogen atom ones in the limit of large
$n$, is accurate whenever the mass in the ground state is much larger
than the combined mass in all the excited states. We have then
analytically derived the mass distribution that is expected among all
the excited states in a steady-state situation of dynamical
relaxation. In this analytical solution, the mass in each excited
state is $m_q \propto n^{-1/3}\propto r^{-1/6}$ (where $r$ is the
characteristic radius of a state of quantum number $n$), leading to
a density profile of a relaxing corona
$\rho \propto N_q m_q / r^3 \propto r^{-5/3}$.

  These relaxing coronae should be reproduced by numerical simulations
of the evolution of a scalar field starting from cosmological initial
conditions, as the ones reported in 
\cite{Schive:2014dra,Schive:2014hza}. Central solitons have indeed been
found to form in these type of simulations, but the presence of the
surrounding coronae, the accretion rates and the accumulated masses of
the solitons predicted by our analytical analysis should be tested. A
study of the mass distribution around the solitons along these lines has
recently been described in \cite{Lin:2018whl}.

  The formation and growth of these solitons will be strongly influence
by the structure of the galaxy that is hosted by the dark matter halo.
Our study will therefore have to be generalized to treat the
accumulation of scalar dark matter on a soliton in realistic halos with
an evolving baryonic mass component. The presence of a central black
hole can also affect the mass of the soliton. Relativistic effects near
the black hole horizon imply a rate of accretion of scalar dark matter
onto the black hole, which might substantially increase the central
black hole masses in some galactic nuclei \cite{Barranco2012}.

  The presence of these axion dark matter solitons may be testable in
the future through a variety of observations, such as detailed mass
modeling in the central region of the Milky Way and several dwarf
galaxies, and gravitational lensing observations.

\acknowledgments
 We are glad to acknowledge useful discussions with Tom Broadhurst,
Nick Kaiser, and Helvi Witek.
This work was supported in part by Spanish grant
AYA2015-71091c and Maria de Maeztu grant MDM-2014-0367 of ICCUB.
J.S. is supported by EU Networks FP10 ITN ELUSIVES (H2020-MSCA-ITN-2015-674896) and INVISIBLES-PLUS
(H2020-MSCA-RISE-2015-690575), by MINECO grant FPA2016-76005-C2-1-P,
research grant 2017-SGR-929, FPA2014-57816-P and PROMETEOII/2014/050.

\bibliographystyle{plain}
\bibliography{paper}

\end{document}